\documentclass[preprint,12pt,number]{elsarticle}
\usepackage{graphicx}
\usepackage{graphics}
\usepackage{tabularx}
\usepackage{makeidx}
\usepackage{makeidx}
\usepackage{epstopdf}
\usepackage{epsfig}
\usepackage{subfig}
\usepackage{colortbl}
\usepackage{colordvi}
\usepackage{verbatim}
\usepackage{amsmath, amsthm}
\usepackage{enumerate}
\usepackage{wrapfig}
\usepackage{hyperref}
\usepackage{dcolumn}
\usepackage{bm}
\usepackage{bbm}
\usepackage{amssymb,mathrsfs}

\journal{Physics Letters B}
\usepackage{bm}% bold math
\def\k{|\vec{k}|}

\def\l{\lambda_*}

\newcommand*{\x}{\mathbf{x}}

\def\beq{\begin{equation}}
\def\eeq{\end{equation}}

\begin{document}

\title{Low Energy Lorentz Violation in Polymer Quantization Revisited}

\author[NK]{Nirmalya Kajuri}
\ead{nirmalya@physics.iitm.ac.in}

\address[NK]{Department of Physics, Indian Institute of Technology Madras, 
Chennai 600036}

\author[GS]{Gopal Sardar}
\ead{gopal1109@iiserkol.ac.in}
\address[GS]{Department of Physical Sciences, Indian Institute of Science 
Education and Research Kolkata, Mohanpur - 741 246, WB, India}

\bibliographystyle{apsrev4-1}

\begin{abstract}
In previous work, it had been shown that polymer quantized scalar field theory 
predicts that even an inertial observer can experience spontaneous excitations. 
This prediction was shown to hold at low energies. However, in these papers it 
was assumed that the polymer scale is constant. But it is possible to relax this 
condition and obtain a larger class of theories where the polymer scale is a 
function of momentum. Does the prediction of low energy Lorentz violation hold 
for all of these theories? In this paper we prove that it does. We also obtain 
the modified rates of radiation for some of these theories.
\end{abstract}
%
%\pacs{04.60.Pp, 03.65.-w}
\maketitle
\section{Introduction.}
The problem of finding the correct quantum theory of gravity is one of the 
biggest challenges in physics today. While the solution to the puzzle remains 
out of reach, many promising approaches have been developed. On one hand there 
are 'top down' approaches like string theory and loop quantum gravity where one 
starts with a theory and tries to obtain experimental predictions. On the other 
 there are 'bottom up' phenomenological approaches where one mainly tries to 
understand the consequence of Planck scale modifications of physics on the 
matter sector. 

Polymer quantized scalar field theory\cite{Hossain:2010eb} is a 
phenomenological model inspired by loop quantum 
gravity\cite{Rovelli:2004tv,thiemann}. Here one first decomposes a free scalar 
field theory into uncoupled harmonic oscillators in momentum space and then 
quantizes each oscillator using polymer quantization\cite{Ashtekar:2002sn} 
which introduces a polymer scale. This procedure yields a modified 
propagator which converges to the standard propagator in the limit of low 
energies. 

Now to test any modified theory we try to find situations where its 
predictions conflict with the predictions of the standard theory, preferably at 
accessible energies. For polymer quantization, the prediction of Unruh Effect 
\cite{Unruh:1976db} (or lack thereof) has proven to be such a scenario where
results obtained from polymer quantization differ significantly from standard
results. Unruh Effect for polymer quantized fields in linearly accelerated frames have been 
studied in \cite{Hossain:2014fma,Hossain:2015xqa}. The case of rotating frames 
have been studied in \cite{Stargen:2017xii}. But perhaps the most striking 
results have been established for inertial frames. 

To understand this result, we 
should first note that polymer quantization violates Lorentz symmetry and 
establishes a preferred frame. It was shown in \cite{Kajuri:2015oza} that a 
detector moving with constant velocity with respect to this frame can detect 
radiation, if it is coupled to a polymer quantized field. Furthermore it was 
found that such detection occurs at low energies. In \cite{Husain:2015tna} it 
was established that there is a critical velocity such that a detector moving 
above this velocity will detect radiation. The rates of radiation were 
calculated in this paper and it was found that they cannot be suppressed by 
increasing the polymer scale. 

However, a restrictive assumption had been made 
while polymer quantizing the scalar field theory in \cite{Hossain:2010eb} It was 
assumed that the polymer scale is a constant. This need not be the case! Recall 
our description of polymer quantization of scalar field. First the field is 
decomposed into harmonic oscillators, one at each point in the space of spatial 
momenta. Then each harmonic oscillator is polymer quantized. As we will see in 
more detail later, this quantization requires the introduction of a scale, which 
we call the polymer scale. In \cite{Hossain:2010eb}, the polymer scale was 
assumed to be the same for all the oscillators. But clearly, this assumption can 
be relaxed. It is a natural extension of \cite{Hossain:2010eb} to consider the 
polymer scale to be a function of $|\vec{k}|$. A running polymer scale is also 
natural from the perspective of renormalization group flow. 

Making this extension, we 
arrive at a large class of polymeric theories, one for each possible 
$\lambda(|\vec{k}|)$. The only stipulation we must put on these theories is that 
they reproduce the standard field theory propagator at the low energy limit. We 
can now ask, do one or more of these theories not violate Lorentz symmetry at 
low energies. This is the question that we address in this paper. Surprisingly, 
we find that \textit{none} of these theories can evade the fate of the original, 
all of them predict that an inertial detector will click at a certain
critical velocity. We obtain a proof of why this should be so. We perform numerical experiments to find out the critical velocities for different theories. Surprisingly, we find that critical velocities turn out to have the same value for very different polymeric theories. Further investigation is necessary to understand why this should be so. Our result 
strengthens the existing result of low energy violation for polymer quantized 
theories. We also obtain the rates of radiations for some of these theories.

The paper is organized as follows. In the following section we recall polymer 
quantization of scalar fields and then modify it by introducing a momentum 
dependent polymer scale. In section III we test some of these theories 
numerically to see if they predict the clicking of an inertial detector and find 
that they do. In section IV we give proof of why this must be so. Section V 
presents our numerical results of the rates of radiation in some of these 
theories. The final section summarizes our results. 
\section{Polymer quantization with variable polymer scale}
In this section we briefly review polymer quantization of scalar 
field\cite{Hossain:2010eb} and then extend it by introducing momentum dependence 
in the polymer scale. First, let's recall polymer quantization of a harmonic 
oscillator\cite{Ashtekar:2002sn}.
In polymer Hilbert space, the position operator $\hat{x}$ and translation 
operator $\hat{U}(\lambda)$ are considered to be basic operators. Since the 
translation operator is not weakly continuous in  the parameter $\lambda$, the 
momentum operator does not exist in polymer Hilbert space. However, one can 
define the momentum operator as 
$\hat{p}_{\lambda}=1/(2i\lambda)(\hat{U}(\lambda)-\hat{U}(-\lambda))$ and one 
can recover usual momentum operator by taking the limit $\lambda 
\rightarrow 0$. In polymer Hilbert space the limit $\lambda \rightarrow 0$ does 
not exist and $\lambda$ is considered as a fundamental scale. By choosing 
 $\lambda$ to be $\lambda_*$, the Hamiltonian of simple harmonic oscillator can 
be expressed as:
\begin{align}
\hat{H} = \frac{1}{8m\lambda_*^2}(2 -\hat{U}(2\lambda_* )-\hat{U}(-2\lambda_* 
)) + \frac{m\omega^2 \hat{x}^2}{2}~.
\end{align}
Note that it is at this step that the polymer scale enters the theory. This 
modifies the Schrodinger equation to:
\begin{align}
\frac{1}{8m\lambda_*^2}( 2- 2\cos(2\lambda_*p))\psi - \frac{m\omega^2 }{2} 
\frac{\partial^2 \psi}{\partial p^2} = E\psi ~.
\end{align}
This can be mapped to a Mathieu equation through the following redefinitions: 
\begin{align}
u = \lambda_*p +\pi/2  \,, \qquad \alpha = 2E/{g\omega} - 1/{2g^2} \,, \qquad g 
=m\omega \lambda_*^2 ~.
\end{align}

With these redefinitions the above equation takes the standard form of the 
Mahtieu equation:
\begin{align}
\psi^{\prime \prime}(u) + (\alpha - \frac{1}{2}g^{-2}\cos(2u))\psi(u) = 0~.
\end{align}
This equation admits periodic solutions for certain values of $\alpha$:
\begin{align}
\psi_{2n}(u) &= \pi^{-1/2} ce_n(1/{4g^2}, u),\qquad \alpha =A_n(1/{4g^2})~,\\
\psi_{2n+1}(u) &= \pi^{-1/2} se_{n+1}(1/{4g^2}, u),\qquad \alpha 
=B_n(1/{4g^2})~,
\end{align}
where $ ce_n, se_n (n=0,1\dots)$ are respectively the elliptic cosine and sine 
functions and $A_n, B_n$ are the Matheiu characteristic value functions. The 
energy eigenvalues of the polymer harmonic oscillator are given by: 
\begin{align}
\label{energy}\frac{E_{2n}}{\omega}=\frac{2g^2A_n(1/4g^2)+1}{4g}~,
\end{align}
\begin{align}
\label{energyy}\frac{E_{2n+1}}{\omega}=\frac{2g^2B_{n+1}(1/4g^2)+1}{4g}~.
\end{align}

Now let us recall polymer quantization of scalar fields. Here the starting point 
is the free Klein Gordon field. First one takes  decomposes this field into 
uncoupled harmonic oscillators with Hamiltonians: 
\begin{align}
H_{\k} =\frac{\pi_{\k}^2}{2} +\frac{{\k}^2\phi_{\k}^2}{2}~.
\end{align}

Now each of these harmonic oscillators can be polymer quantized by introducing 
some polymer scale $\lambda_*$. In \cite{Hossain:2010eb} each of these 
oscillators were quantized using the \textit{same} polymer scale. This gives the 
polymer Wightman function: 
\begin{align}
\label{Wightman}\nonumber &\langle 0|\hat{\phi}(t,\x) 
\hat{\phi}(t',\x')|0\rangle =\\ & \sum_{n=0}^{\infty} \int 
\frac{d^3\vec{k}}{(2\pi)^3}e^{i 
{\vec{k}}\cdot(\x-\x')}|c_{4n+3}|^2 e^{-i\Delta E_{4n+3} (t-t')} ~,
\end{align}
where 
\begin{align}
\label{energyyy}\Delta E_n \equiv E_n(g) - E_0(g)~,
\end{align}
and
$c_n(g) = \langle n_{}| \hat{\phi}_{\k} |0_{\k}\rangle$
and $g = \lambda_*^2|\vec{k}| $ .

Using the asymptotic expansions for Mathieu value functions, one can obtain the 
propagator for low momenta ($g \ll 1$): 
\begin{align}
D_p = \frac{i(1-2\lambda_*^2|\vec{k}|)}{p^2-\lambda_*^2|\vec{k}|^3-i\epsilon}~.
\end{align}
This can be seen to go to the usual limit as $g \rightarrow 0$. This completes 
the review of standard polymer quantization. Now we note that all the 
oscillators need not be polymer quantized using the same polymer scale 
$\lambda_*$. Oscillators corresponding to different momenta can have different 
polymer scales \footnote{We note that another possible extension of polymer 
quantization could come from making the energy spacings field dependent. In this 
case the oscillators won't be governed by Mathieu equations. This would be an 
interesting avenue to pursue in future.}. In other words, the polymer scale can 
be a function of momenta. 
In particular, since only $|\vec{k}|$ enters the oscillator Hamiltonian, the 
polymer scale should be taken to be a function $\mu(|\vec{k}|)$. With this 
modification we now have a large class of polymeric theories, one for each 
possible $\mu(|\vec{k}|)$. The new formula for the Wightman function and 
propagator will have the same form as above, with the only modification that 
constant $\lambda_*$ will be replaced by $\mu(|\vec{k}|)$ wherever it appears. 
So far we have not imposed any restrictions on $\mu(|\vec{k}|)$. We will now 
demand that it reduces to the standard field theory propagator at the limit of 
low momenta. The modified polymer propagator at low energy is given by: 
\begin{align}
D_p^\mu = 
\frac{i(1-2|\vec{k}|\mu(|\vec{k}|)^2))}{p^2-|\vec{k}|^3\mu(|\vec{k}
|)^2)-i\epsilon}~.
\end{align}
For this to reduce to the standard propagator we must have $\mu(|\vec{k}|)^2\k 
\rightarrow 0$ in the low momentum limit. Thus we have our only condition on 
$\mu(\k)$:
\begin{align}
\label{condition}\mu(|\vec{k}|)^2 \k \rightarrow 0 \,~\text{when}~|\vec{k}| 
\rightarrow 0 ~.
\end{align}
\section{Low energy Lorentz violation in extended polymeric theories}

\begin{figure}[!htb]
\vskip 15pt
\begin{center}
\resizebox{230pt}{160pt}{\includegraphics{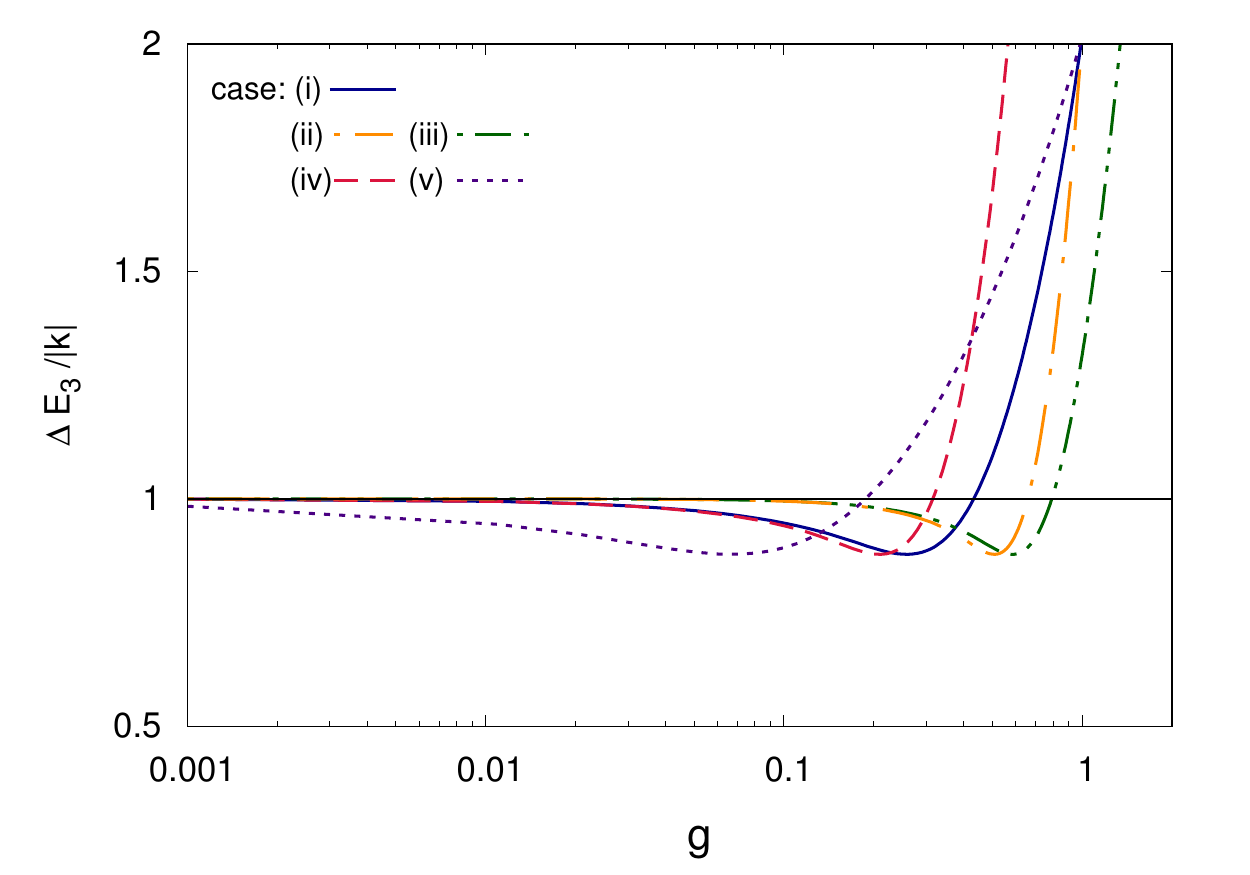}}

\caption{Plot of $ \Delta E_{3}^\mu/|\vec{k}|$ with $g=\l^2\k$. The solid blue 
line represents the standard polymer case and other dashed lines represent for 
different functions  $\mu(\k)$. }
\label{fig:1-std}
\end{center}
\end{figure}

In this section we investigate whether one or more of these theories can avoid 
violating Lorentz symmetry at low energies. First let us recall the criterion 
for an inertial detector to click given in \cite{Husain:2015tna}. In 
\cite{Husain:2015tna} it was shown that the rate of radiation for an inertial 
detector with energy gap $\Omega$ coupled to a polymer scalar field is given by:
\begin{align}\label{resopnsefunction}
F(\Omega) = \nonumber &\frac{1}{2 \pi  \sinh \beta}\sum_{n=0}^{\infty} \int d\k 
~
~\k~|c_{4n+3}|^2\\&  \theta \left(\l^2\k\sinh \beta-|\l^2\Omega +\l^2\k 
\frac{\Delta E_{4n+3}(\lambda_{*}^2\k)}{|\vec{k}|}\cosh \beta|\right) ~.
\end{align}
where $\beta$ is the rapidity of the detector with respect to the preferred 
frame. From the above expression one can see that the rate will vanish if 
$\frac{\Delta E_{4n+3}}{|\vec{k}|} \geq 1$ for all $|\vec{k}|$ and all $n$. But 
if $\frac{\Delta E_{4n+3}}{|\vec{k}|}$ dips below 1 for any range of $|\vec{k}|$ 
and for any $n$, we will have an inertial detector registering radiation. In 
\cite{Husain:2015tna} it was shown that for the standard polymer theory 
$\frac{\Delta 
E_{4n+3}}{|\vec{k}|}$ never dips below 1 for $n > 0$. But $\frac{\Delta 
E_{3}}{|\vec{k}|}$ does dip below 1. Now for the extended class of polymeric 
theories that we have introduced, the above expression for rate holds with the 
replacement of constant $\lambda_*$ by $\mu(\k)$ wherever the former occurs. 
Thus the criterion for an inertial detector not clicking for these theories is $ 
\frac{\Delta E_{4n+3}^\mu}{|\vec{k}|}\geq 1$ for all $\k$. Here $\Delta 
E_{4n+3}^\mu$ is obtained by replacing $\l$ by $\mu(\k)$ in the equations 
\eqref{energy}, \eqref{energyy} and \eqref{energyyy}. We will prove that for 
any 
$\mu(\k)$ that satisfies \eqref{condition} $ \frac{\Delta E_{3}^\mu}{|\vec{k}|}$ 
must dip below unity. We will also show that the dip below unity for all these 
theories occurs at low momenta. Thus all these theories exhibit low energy 
Lorentz violation. Before going into the proof let us pause to look at some 
numerical evidence for this claim. In FIG.(\ref{fig:1-std}) we have plotted $ 
\frac{\Delta 
E_{3}^\mu}{|\vec{k}|}$ for different functions $\mu(\k)$. We have considered the following functions  $\mu(\k)$:

Case:(i) $\mu^2(\k)=\l^2$ (the standard polymer scale case) \\
(ii) $\mu^2(\k)= \l^4 \k $
(iii)  $\mu^2(\k)=\l^2\left(1-e^{-\l^2\k}\right)$ \\ 
(iv) $\mu^2(\k)=\l^2 e^{\l^2\k}$ and 
(v) $\mu^2(\k) = \frac{\l}{\sqrt{\k}}$ .
We see that for all of 
them, there is a dip below unity at low momentum.
Now we proceed to the proof. The first step is to write $ \frac{\Delta 
E_{3}^\mu}{|\vec{k}|}$ in terms of Mathieu characteristic value functions:
\begin{align}
\label{mathieu}
\frac{\Delta E_{3}^\mu}{|\vec{k}|} = 
\frac{\mu^2(\k)\k}{2}\left(B_2(\frac{1}{4\mu^4(\k)\k^2}) - 
A_0(\frac{1}{4\mu^4(\k)\k^2})\right) ~.
\end{align}

Next we note from condition \eqref{condition} that as $k\rightarrow 0$, 
$\frac{1}{4\mu^4(\k)\k^2}\rightarrow \infty$. This allows us to use the 
asymptotic expansion of Mathieu characteristic value functions at low momenta. 
Using 
the asymptotic expansion we get that:
\begin{align}
\nonumber B_2(\frac{1}{4\mu^4(\k)\k^2})-&A_0(\frac{1}{4\mu^4(\k)\k^2})\\& 
=2\frac{1}{\k \mu^2(\k)}-\k\mu^2(\k)~.
\end{align}

Combining this with \eqref{mathieu} we obtain:
\begin{align}
\nonumber\frac{\Delta E_{3}^\mu}{|\vec{k}|}& = 
\frac{\mu^2(\k)\k}{2}\left(2\frac{1}{\k 
\mu^2(\k)}-\k\mu^2(\k)\right)\\&\nonumber =1-\frac{\k\mu^2(\k)}{2}\\&= <1~.
\end{align}

This proves that all theories with variable polymer scales that satisfy the 
consistency condition \eqref{condition} will exhibit low energy Lorentz 
violation.
But can we make a statement about the magnitude of critical velocity from this? In (Husain and Louko) the critical velocity had been found to be well within experimental reach. However for general polymeric theories it is possible that the critical speed is high enough to avoid detection by existing experiments.

 Unfortunately, the asymptotic analysis we employed cannot determine the critical velocity and we must resort to numerical experiments. as For a given polymeric theory, the critical velocity needs to be determined numerically. In the following section we consider several examples of polymeric theories and test them numerically to obtain the critical velocity.

\section{Rates of radiation for different polymeric theories}

In this section we present our numerical estimation of  critical velocities and radiation rates for 
different polymeric theories with different $\mu(\k)$ s.
%%%%%%
Firstly, surprisingly numerical estimation shows that the minimum value of 
$\Delta 
E_3/\k\approx 0.8781$ is same for all polymeric theories which are considered 
here. From the Eq.(\ref{resopnsefunction}) one can see that the critical 
rapidity (for 
$\Omega>0$, above which rapidity the detector's excitation rate is non-zero) 
depends only on the $\Delta E_3/\k$. Thus the critical rapidity 
$\beta_c=arctanh((\Delta E_3/\k)_{min})\approx 1.3675$ remains same for all 
these polymeric theories. 
We have plotted the rate of radiation $F(\Omega)$ with the rapidity $\beta$. We 
find that, for a given value of $h=\l^2 \Omega$, the detector starts clicking 
at different rapidity in different polymeric theories. This is due to the fact 
that the occurrence of minima of $\Delta E_3/\k$  at different value of $g$ in 
different polymeric theories though it has same value for all polymeric 
theories. Secondly, the rates of radiations vary somewhat, but are not 
suppressed 
compared to the standard case with constant polymer scale. 
%%%%%%%

Our numerical experiments suggest that violation at low velocity is a robust feature of polymeric theories. Further investigation is required to provide an explanation for the critical velocity being same in all cases.

In all the figures the three lines denote different values of the parameter 
$h$. The red (the highest pick), blue and green (the lowest pick) 
lines denote $h=0.01,0.05 $ and $0.1$ respectively.

\begin{figure}[!htb]
\vskip 15pt
\begin{center}
\resizebox{200pt}{150pt}{\includegraphics{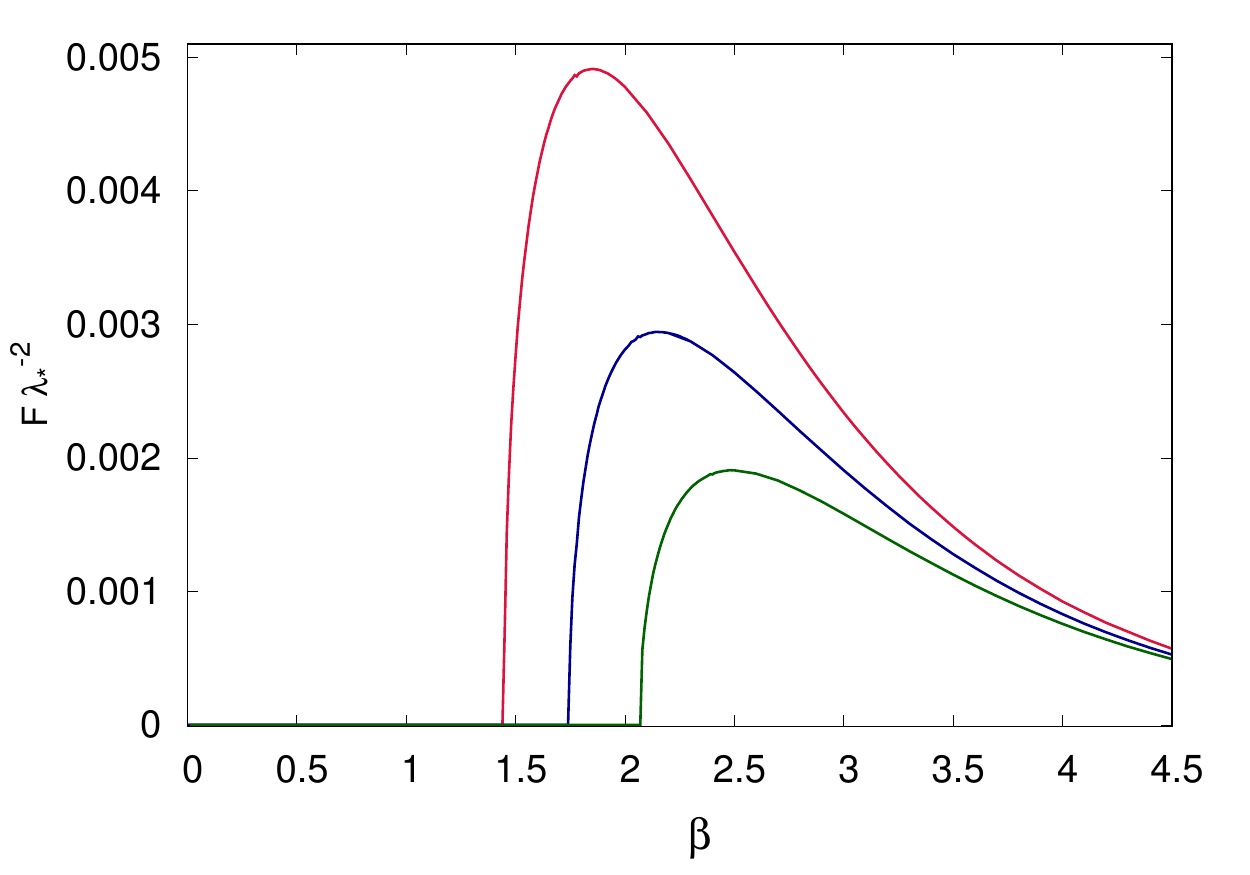}}
\caption{Radiation rate for the standard case with constant polymer scale.}
\label{fig:2-std}
\end{center}
\end{figure}

 \begin{figure}[!htb]
  \centering
  \includegraphics[width=9.1 cm]{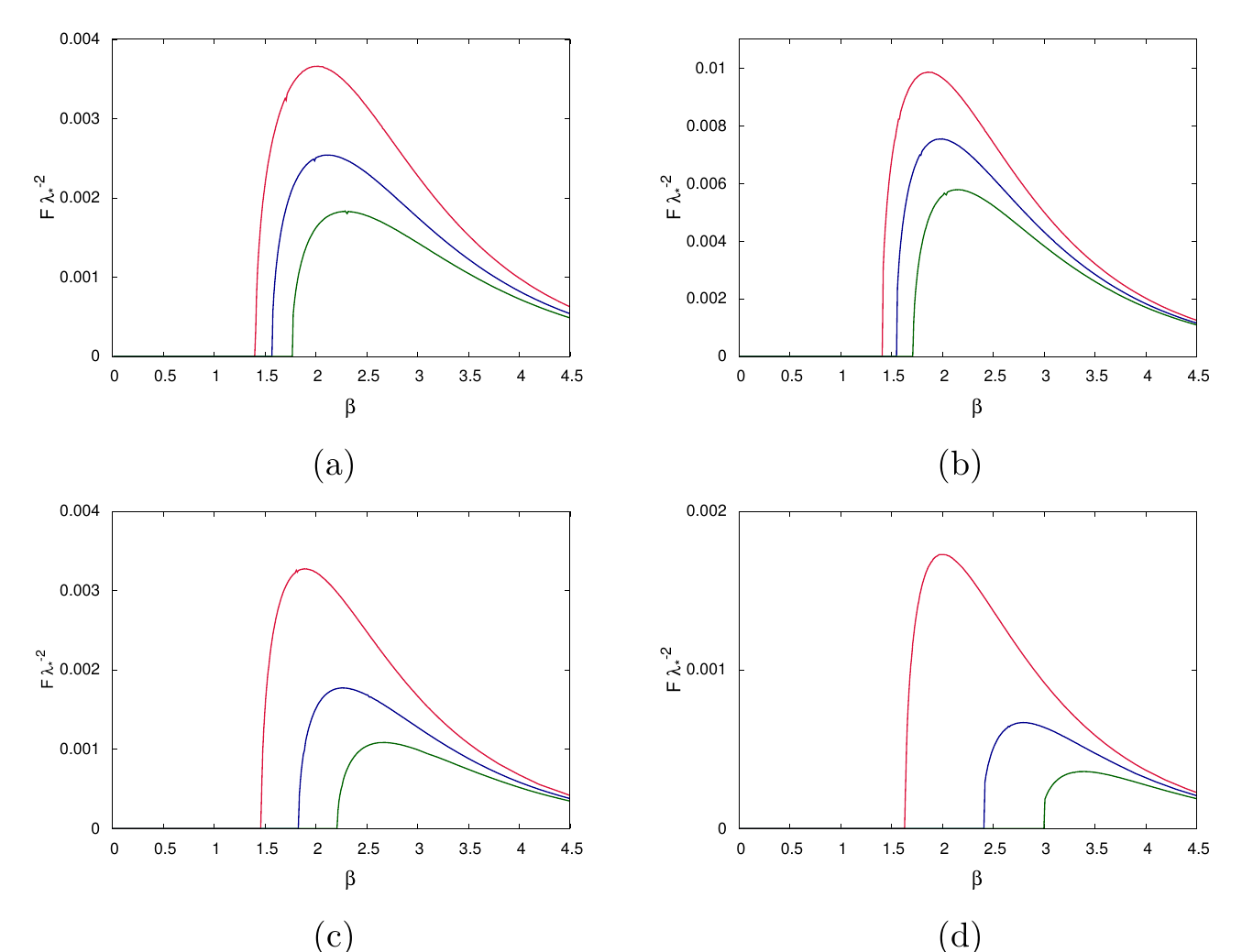}
  \caption{ The radiation rate for different functions $\mu(\k)$. The 
sub-figures (a), (b), (c) and (d) represent the case (ii), (iii), 
(iv) and (v) respectively. }
\end{figure}

\section{Summary}

In this paper we first extended polymer scalar field theory into a larger class 
of theories by allowing the polymer scale to vary with momentum. Then we 
investigated whether all of these theories also exhibit low energy Lorentz 
violation. We gave a proof that they all do indeed violate Lorentz symmetry at 
low energies (as demonstrated by the clicking of an inertial detector travelling 
at low velocities). We also verified this with numerical evidence for several 
possible cases.
We also found through numerical experiments that the critical rapidity $\beta_c=1.3675$ remains same for several possible polymeric theories. This is an intriguing feature which requires further investigation. Finally, we obtained the rates of radiation for different polymeric theories.

It should be possible to test our predictions at the Relativistic  Heavy Ion  
Collider (RHIC). The dipole moment interaction between atoms and 
electromagnetic fields closely resembles the Unruh-DeWitt detector. Since the 
critical rapidity is much below than $\beta \approx3$  
attained by ions at RHIC \cite{rhic}, polymeric theories should be 
verifiable at RHIC. 

Our result thus strengthens previous work on low energy Lorentz violation in 
polymer quantized theories, closing a possible loophole.

\section{Acknowledgements} 
\noindent
N.K would like to thank D. Jaffino Stargen for valuable discussions and G.S 
would like to thank Golam M Hossain for many discussions over the years that 
helped to deepen his understanding of polymer quantization. We would like to 
thank Jorma Louko for his valuable comments on the manuscript.

\end{document}